\documentclass[12pt]{article}
\usepackage[dvips]{graphicx,color}

\newcommand{\msbar}{\mbox{\small {$\overline {MS}$}}}

\newenvironment{Mylis}[1][]
{\begin{list}
 {#1}
 {\settowidth{\labelwidth}{#1}\setlength{\labelsep}{0.5em}
  \setlength{\leftmargin}{\labelwidth}\addtolength{\leftmargin}{\labelsep}
  \setlength{\rightmargin}{0em}
  \setlength{\parsep}{0.3\parskip}\setlength{\itemsep}{0.3\parskip}
  \setlength{\topsep}{0ex}}
}
{\end{list}}

\newcommand{\mylis}[2][]
{\begin{Mylis}[#1] #2 \end{Mylis}}

\newcommand{\wXsecErr}{
\begin{figure}[ht]
\vspace{-3 ex} \centerline{
\resizebox{0.7\textwidth}{!}{\includegraphics{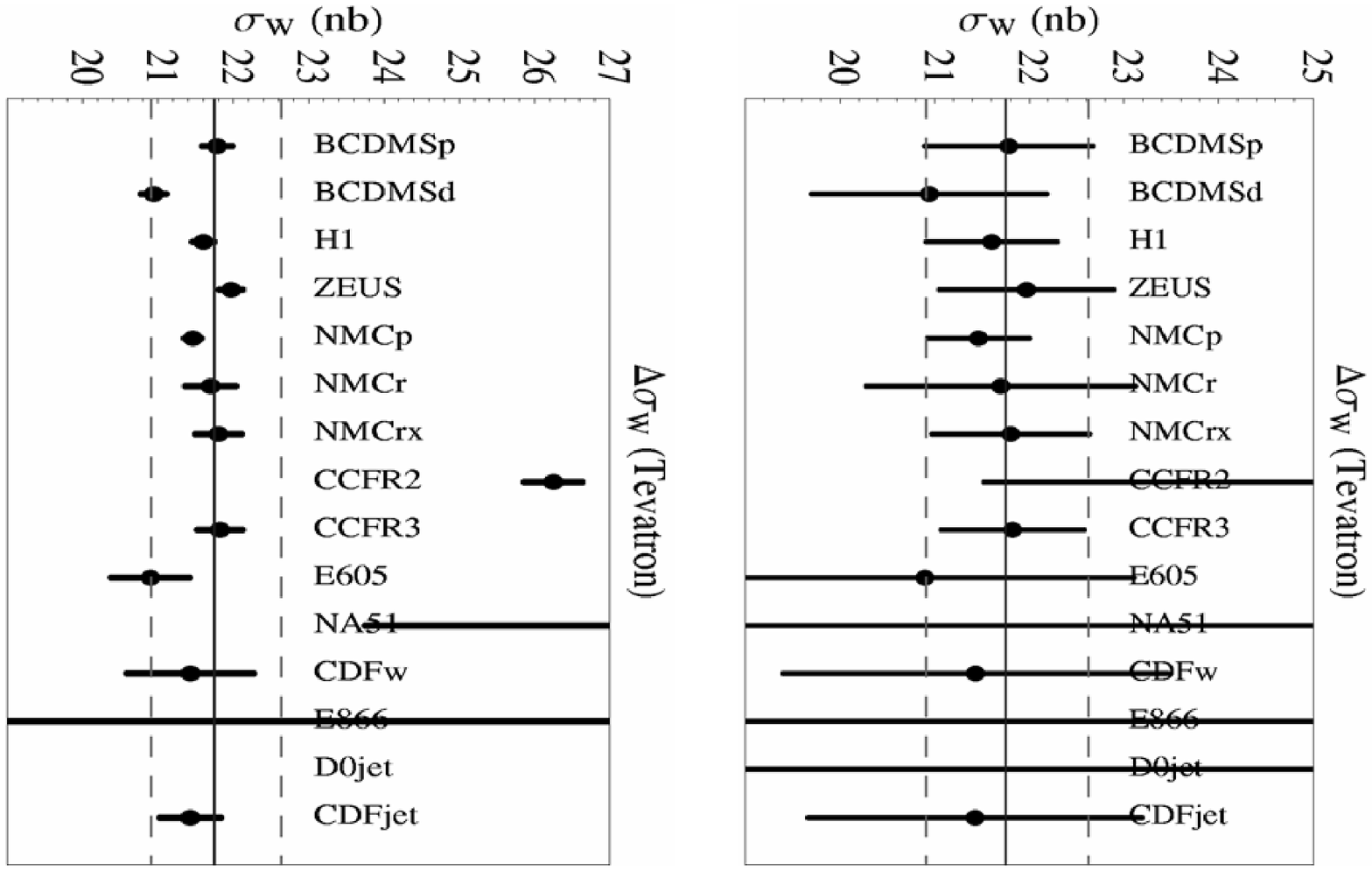}} }
\caption{Predicted value of $\sigma_W$ at the Tevatron: (a) $\Delta\chi^2=1$
error ranges for individual experimental data sets evaluated from PDF sets
obtained by the Lagrange Multiplier method in constrained global fits; and
(b) 90\% confidence level ranges for the same data sets and PDF sets.}
\label{fig:wXsecErr} \vspace{-4ex}
\end{figure}
}
\newcommand{\Unc}{
\begin{figure}[ht]
\vspace{-3ex}\hfill
\resizebox{0.42\textwidth}{!}{\includegraphics{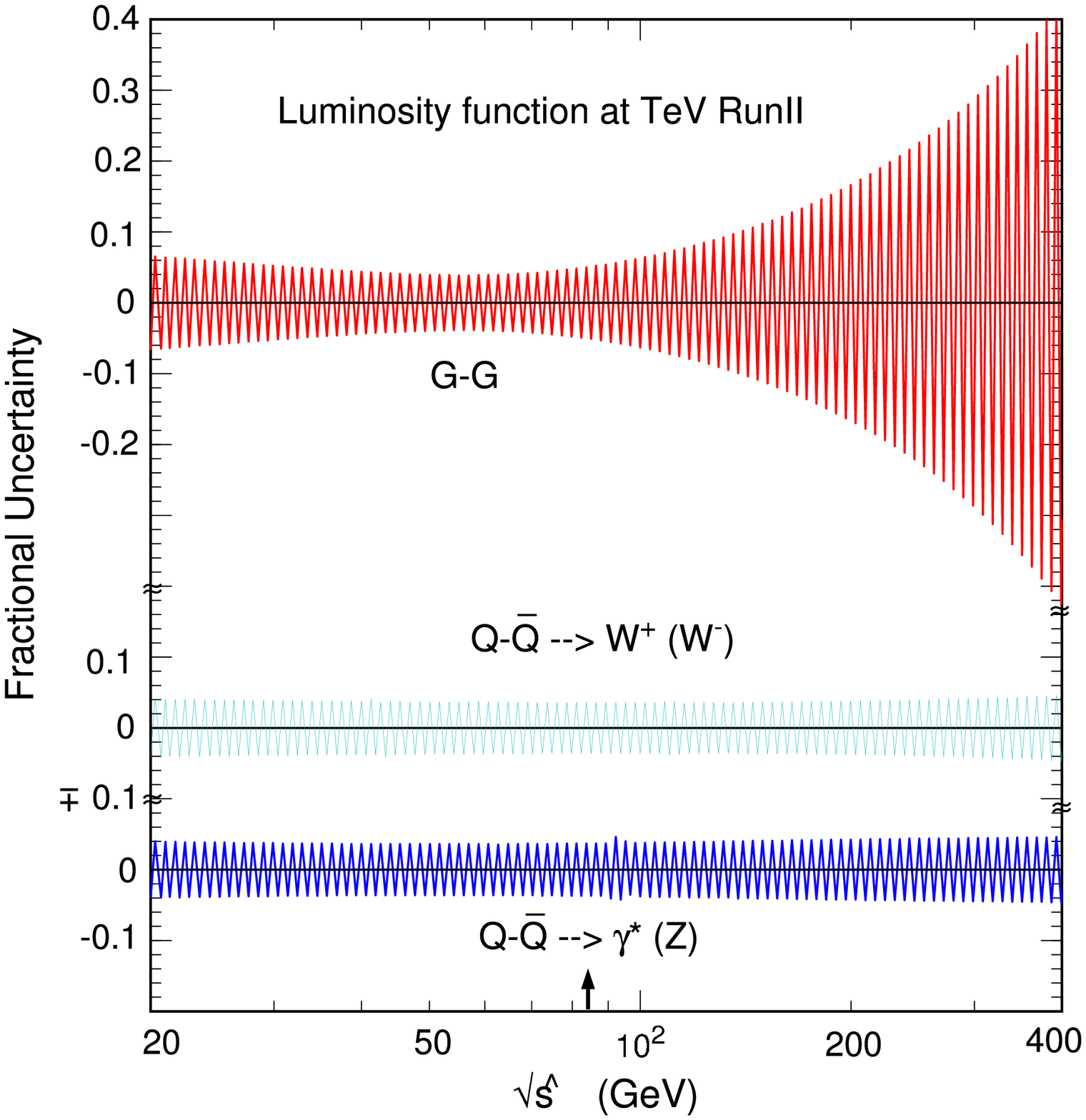}} \hspace{3em}
\raisebox{2ex}{
\resizebox{0.40\textwidth}{!}{\includegraphics{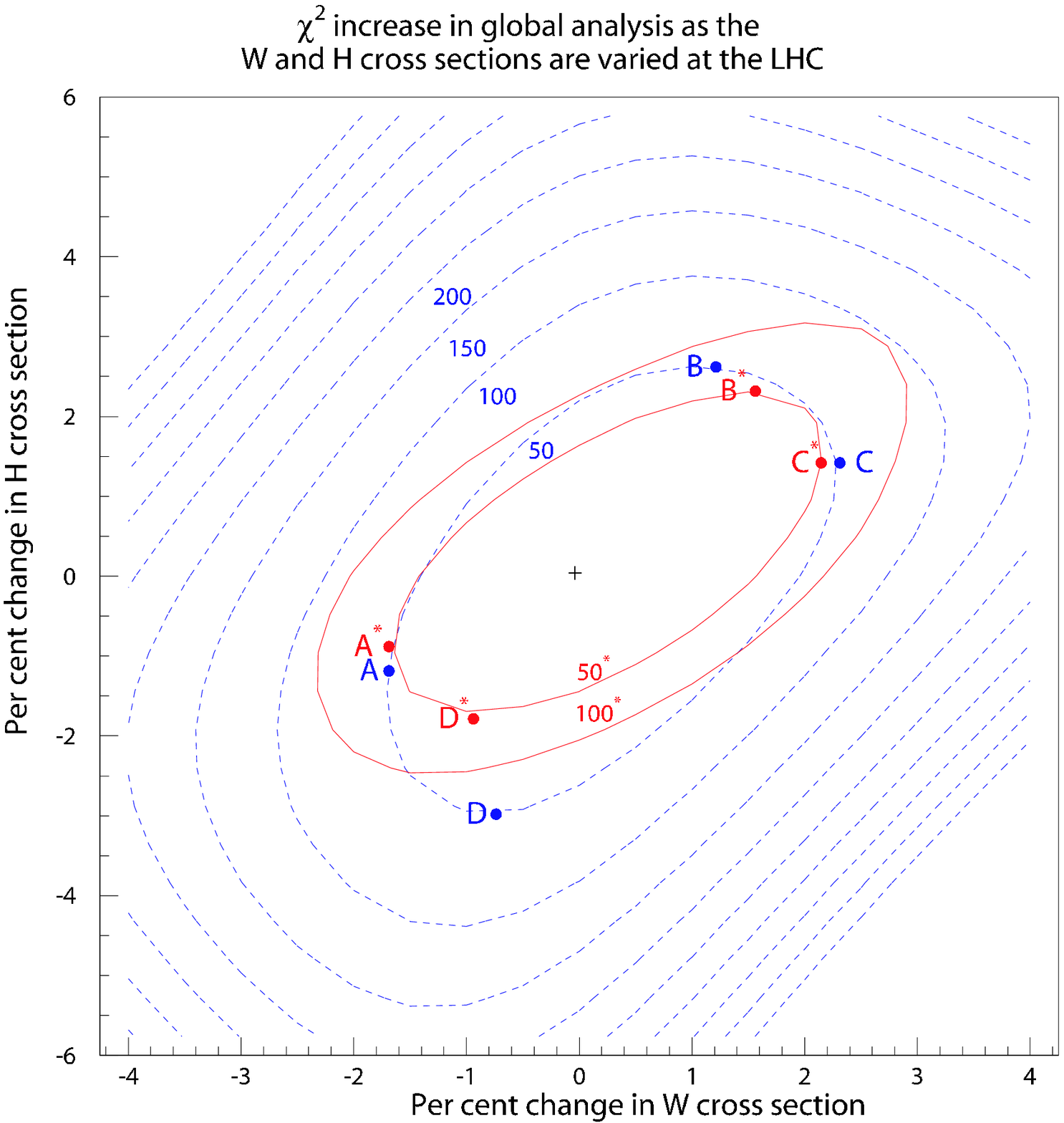}}
\hfill}
\caption{Examples of ranges of predictions at Tevatron Run II and LHC
associated with uncertainties of PDFs due to experimental input. See text.}
\label{fig:Unc} \vspace{-3ex}
\end{figure}
}
\newcommand{\mrst}
{\begin{figure}[ht]
\parbox[3in]{\textwidth}
{\hfill
\raisebox{3ex}{\resizebox{0.43\textwidth}{!}{\includegraphics
[clip=]
{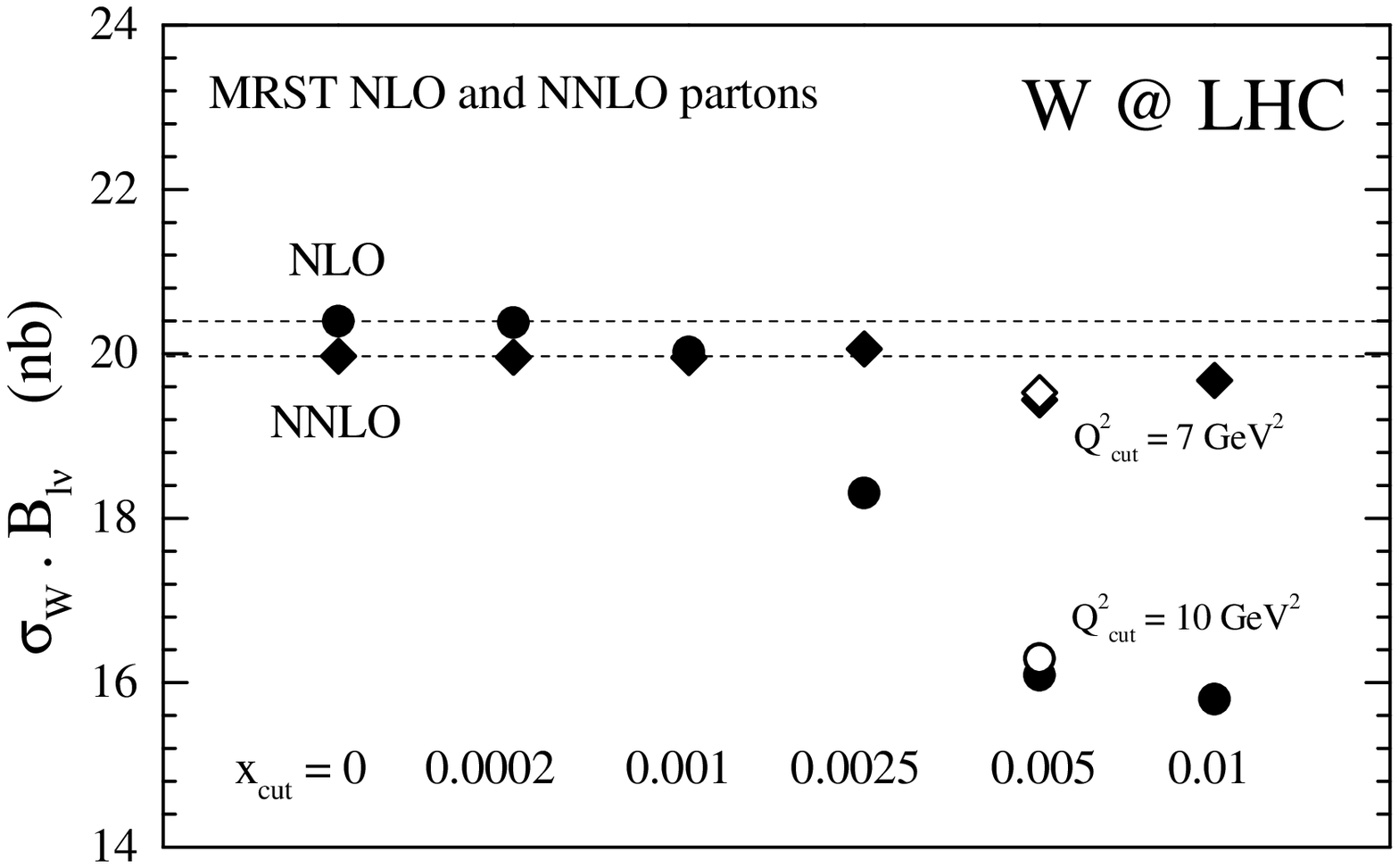}}}
\hspace{3em}
\resizebox{0.41\textwidth}{!}{\includegraphics
[clip=]
{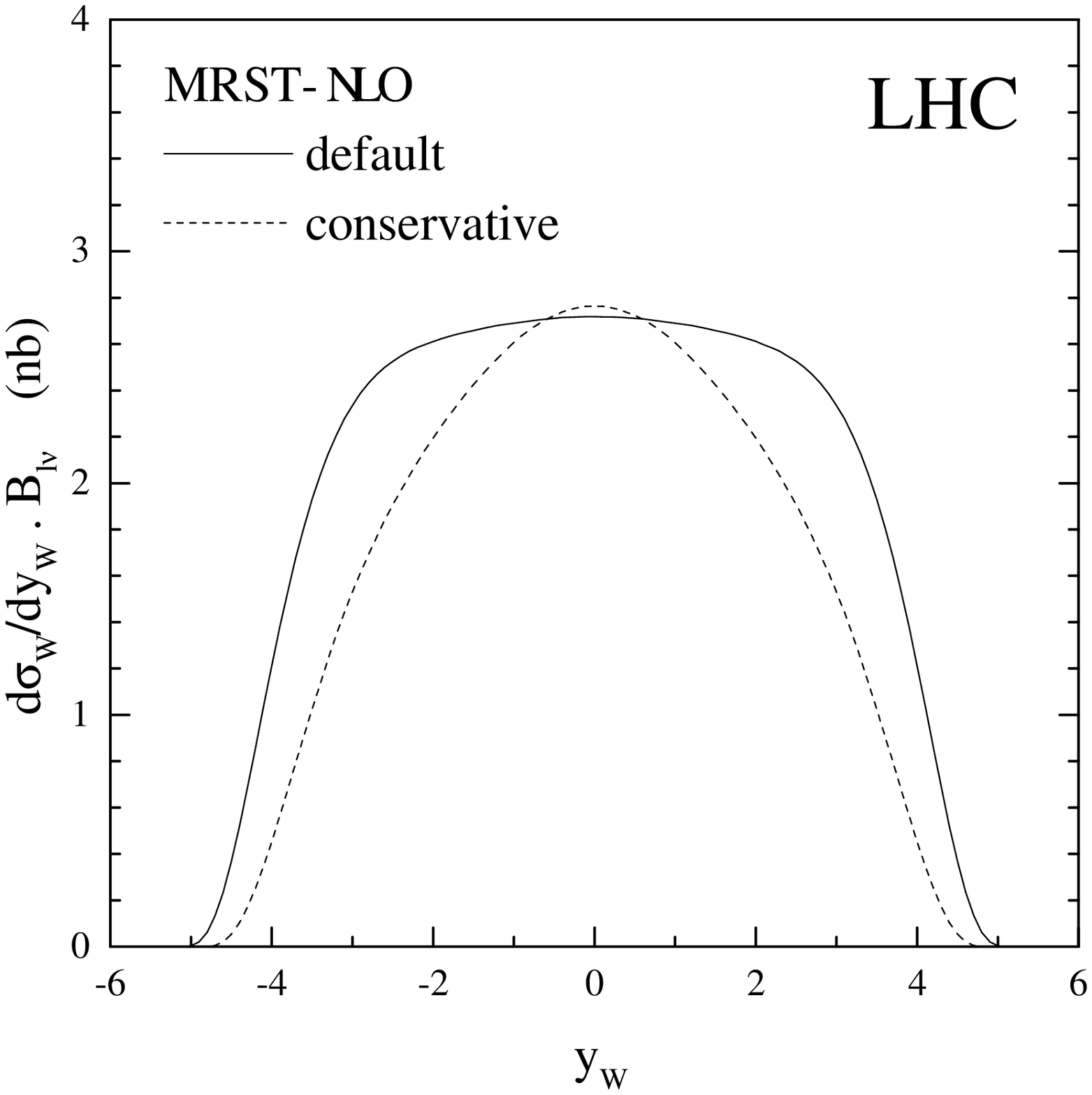}
\hfill}
\caption{(a) Dependence of MRST predictions on W total cross section at LHC
on kinematic cuts of input data; (b) W rapidity distribution according to the
MRST default and the ``conservative" (relatively high $x$ and $Q$ cuts)
PDFs.}}
\label{fig:mrst}
\end{figure}
}
\newcommand{\cteq}
{\begin{figure}[ht] \hfill \raisebox{2ex}{
\resizebox{0.42\textwidth}{!}{\includegraphics [clip=] {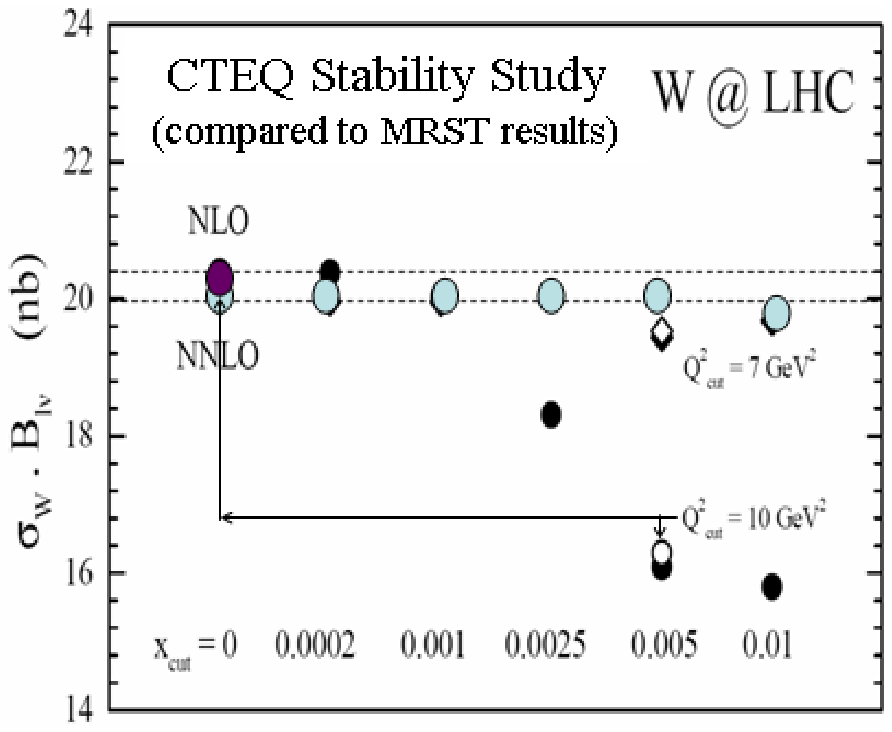}}
} \hspace{3em} \resizebox{0.42\textwidth}{28 ex}{\includegraphics [clip=]
{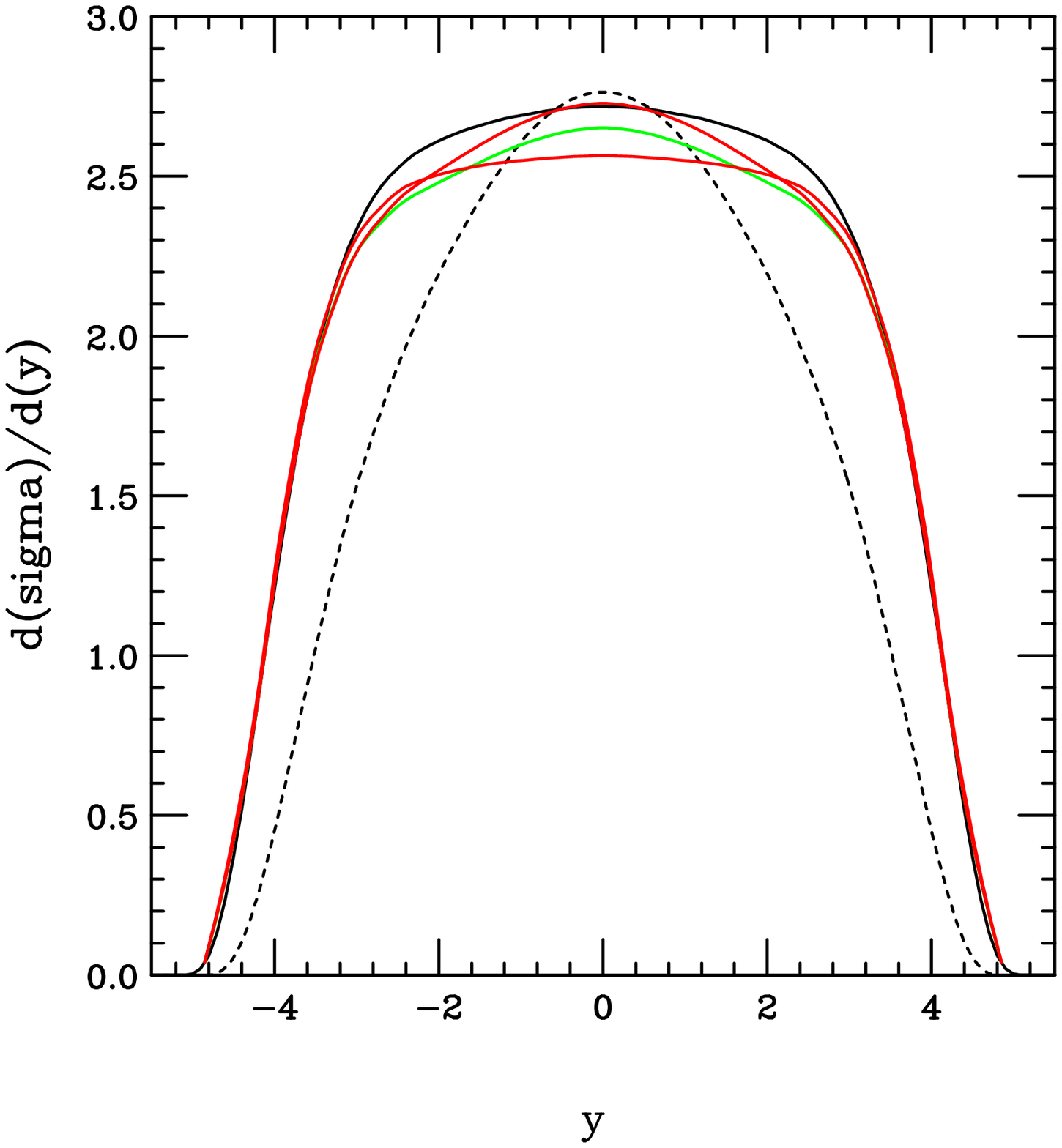} \hfill} \caption{(a) Dependence of CTEQ predictions
(light dots) on W total cross section at LHC on kinematic cuts of input data,
super-imposed on Fig.~\protect\ref{fig:mrst}a; (b) W rapidity distribution at
LHC, CTEQ6 prediction plus those extremized w.r.t.\ the 2nd moment of y in
solid color lines, compared to that of the MRST default and conservative PDF
predictions (in black solid and dotted lines).} \vspace{-3ex}
\label{fig:cteq}
\end{figure}
}
\newcommand{\parabolas}
{\begin{figure}[ht] \hfill \resizebox{0.41\textwidth}{25
ex}{\includegraphics[clip=] {figs/LM+glu.eps}} \hspace{3em}
\resizebox{0.41\textwidth}{25 ex}{\includegraphics[clip=] {figs/LM-glu.eps}
\hfill } \caption{Ranges of uncertainty on the ($W^+ + W^-$) production cross
section at the LHC: $\Delta\chi^2$ (over best fit) vs.\ $\sigma_W$ for
several choices of kinematic cuts to the input data. (a) Normal fits with
positive PDFs; and (b) Similar fits, but allowing the input gluon
distribution at $Q=1.3$ GeV to go negative at small $x$. Note the stability
of the minima of the curves---reflecting the same results of
Fig.\,\ref{fig:cteq}a. } \vspace{-3ex} \label{fig:parabolas}
\end{figure}
}

\begin{document}

\title{Global QCD Analysis and Hadron Collider Physics%
\footnote{Contribution to the Proceedings of \emph{The 15th International
Topical Conference on Hadron Collider Physics}, HCP2004, Michigan State
University, June, 2004.}
}
\author{Wu-Ki Tung%
\footnote{This review incorporates results of some recent, yet unpublished,
work done in collaboration with
Joey Huston, Jonathan Pumplin and Daniel Stump at Michigan State University.}%
\\Michigan State University, E. Lansing, MI, USA}

\date{}

\maketitle

\begin{abstract}
The role of global QCD analysis of parton distribution functions (PDFs) in
collider physics at the Tevatron and LHC is surveyed. Current status of PDF
analyses are reviewed, emphasizing the uncertainties and the open issues. The
stability of NLO QCD global analysis and its prediction on ``standard candle"
W/Z cross sections at hadron colliders are investigated. The importance of the
precise measurement of various W/Z cross sections at the Tevatron in
advancing our knowledge of PDFs, hence in enhancing the capabilities of
making significant progress in W mass and top quark parameter measurements,
as well as the discovery potentials of Higgs and New Physics at the Tevatron
and LHC, is emphasized.
\end{abstract}

\vspace{-2ex}
\subsection*{Introduction}

As the physics program at the Tevatron advances from Run I to Run II, and as
the planning of LHC physics shifts into high gear, it is important to assess
the foundation on which most of the relevant physics calculations are based.
All experimental measurements are made on lepton- and hadron- initial (and
final) states. On the other hand, the fundamental physics we are trying to
unravel is formulated in terms of interactions between leptons, vector
bosons, quarks, gluons, Higgs and other new particles. To make any progress,
we need to ask: How well do we know the parton (quark and gluon) distribution
functions (PDFs) of the nucleon? What important pieces of information on PDFs
are still missing? What are the uncertainties of the ``known" pieces? What
information gained at the Tevatron Run II and HERA II will be important to
predict, and help sort out, the physics at LHC? Some, but not all, of these
questions are addressed in this talk. Due to space limitations, only
representative results from the talk can be included in this written report.
For the same reason, references will be limited to the essential ones.
Details on many topics can be found in recent literature and in related
articles in these proceedings.

\vspace{-2ex}
\subsection*{Status and Open Issues in Global QCD Analysis}

A great deal of progress has been made in global QCD analysis over the last
20 years, due to new experimental input from a variety of hard processes
(DIS, DY, W/Z and jet production in colliders), advances in theoretical
calculations (NLO, NNLO and resummation), and more powerful phenomenological
analysis tools (improved Hessian methods, Lagrange Multiplier technique, and
functional approaches). Nonetheless, many features of the PDFs are still
uncertain, or unknown. These open issues are summarized here, with some brief
background information in each case.

\vspace{1ex}\noindent\textbf{The $u$ and $d$ distributions}\vspace{0.5 ex}%
\\Improved data have caused considerable changes in the PDFs $u(x,Q)$ and
$d(x,Q)$ over the years, particularly at small $x$ (as a direct consequence
of the advent of the HERA data since the mid-1990's). Combined high precision
fixed-target and HERA data, supplemented by constraints from DY and W
production data, now cover roughly the $x$ range of ($10^{-5},0.75$). These
data result in rather well-determined $u$ and $d$ distributions in most
recent global analyses.\ \cite{Mrst23,cteq6} The remaining uncertainties
concern mainly the large $x$ behavior, beyond the measured range,
particularly the $d/u$ ratio. The discrimination between these two flavors is
currently hampered by unknown nuclear corrections associated with the use of
DIS and DY data on deuteron targets. Recent preliminary DY cross section data
from NuSea \cite{NuSea} have stimulated renewed interest in this problem.
However, no conclusion can be reached until the data is finalized. In
principle, more precise $e^{\pm }p$ charged-current scattering data at HERA
($W^{+}/W^{-}$ exchanges) would be able to determine the $d/u$ ratio free of
nuclear effects. So will data, eventually, on the rapidity distributions of
$W^{+}/W^{-}$ production at the LHC.

\vspace{1ex}\noindent \textbf{The gluon distribution}\vspace{0.5 ex}%
\\It is well known that the constraints on $g(x,Q)$ are much looser than for $%
u $ and $d$. Hence, the first determinations of the gluon distribution varied
over a wide range. The initial HERA data forced a much steeper rise of
$g(x,Q)$ toward small $x$. However, more recent global analyses all have a
much more moderate rise, or, in some cases, even a fall, in the small-$x$
behavior of the gluon at low $Q$. An important contributing factor for this
turn-around is the indirect effect of including single-jet inclusive
production data from the Tevatron. These data favor a larger $g(x,Q)$ at high
$x$, which takes away gluons at small $x$ because of the overall momentum sum
rule. The range over which the gluon distribution has developed over these
years shows vividly both how global QCD analysis has been evolving, and how
much further we need to go to determine the parton structure of the nucleon
with confidence. The fractional uncertainty on $g(x,Q)$ is largest at high $%
x $, where experimental constraints are scarce. At small $x$, there is a
whole range of open theoretical and phenomenological issues. In fact, in the
MRST analyses, the necessity of introducing a negative gluon distribution at
low $x$ and small $Q$ has been proposed \cite{Mrst23}. These issues will be
discussed in a later section.

\vspace{1ex}\noindent \textbf{The strangeness sector }\vspace{0.5 ex}%
\\An important, but so far poorly determined, frontier of PDF study is the strangeness
sector. This can be separated into two parts: the total strangeness sea, $%
s_{+}(x)=s(x)+\bar{s}(x)$, and the strangeness charge asymmetry $%
s_{-}(x)=s(x)-\bar{s}(x)$. Advance on both fronts is expected by detailed
study of the recent CCFR-NuTeV dimuon production data in neutrino and
anti-neutrino scattering \cite{CN} in global analysis.\footnote{%
The global analysis context is important: PDFs in PQCD, by definition, are
\emph{universal}, i.e.~they must be applicable to all processes. Specialized
PDFs derived from the analysis of very limited data sets (or processes) that
do not fit a broad range of data may not correspond to true PDFs of nature.}

\vspace{1ex}\emph{Flavor SU(3) breaking:}%
\ \ The naive expectation of flavor SU(3) symmetry,
$s=\bar{s}=\bar{d}=\bar{u}$, assumed in some early PDF studies, is clearly
unrealistic: the strange quark mass alone would induce a difference between
($s,\bar{s}$) and, say, $(\bar{u}+\bar{d})/2$. Experimental evidence suggests
that the ratio $R_{s+}\equiv \frac{s+s}{\bar{u}+\bar{d}}$, is of the order
0.5 at a scale of 1-2 GeV. Up to now, this ratio has mainly been enforced as a
(constant) factor in global analyses, rather than as the result of actual
fitting to data, because the relevant data have not been presented in a form
suitable for global analysis.

\vspace{1ex}\emph{Strangeness Charge Asymmetry:}%
\ \ In order to separate the $s$ and $\bar{s}$ distributions, and hence allow the
measurement of the strangeness charge asymmetry, $s_{-}(x)$, the separate
measurement of $\nu $ and $\bar{\nu}$ cross sections is required. Although
the strangeness number sum rule requires $\int_{0}^{1}s_{-}(x)~dx=0,$ a
charge asymmetric strange sea,
manifested by, e.g.\ a non-zero first moment $[S_{-}]=%
\int_{0}^{1}x~s_{-}(x)~dx,$ has important physical consequences such as the
NuTeV anomaly \cite{NuTeV}.

Quantitative study of this issue is currently limited by the fact that
CCFR-NuTeV measures $\nu $ and $\bar{\nu}$ cross sections with specific
kinematic cuts \cite{CN}, rather than inclusive cross sections that theory
can calculate. In order to perform a global fit incorporating these new data,
a LO model bridging the inclusive and the measured cross sections has
to be invoked. With this caveat, CTEQ has found that (i) a positive value for $%
[S_{-}]$ on the order of 0.0017 is favored; and (ii) the range of
uncertainty, both theoretical and experimental, is rather large:
$-0.001<[S_{-}]<0.004$.\,\cite{StrAsym} The analysis by CCFR-NuTeV, using
their own data only, reached a somewhat different conclusion \cite{CN}. The
two groups are working together to understand the differences, and to extend
both analyses to a full NLO one.

\vspace{1ex}\noindent \textbf{Possible Isospin Violation in the Parton
Structure of the Nucleon}\vspace{0.5 ex}%
\\Interest in possible explanations of the NuTeV anomaly has also
motivated the study of the possibility that $u_{\mathrm{proton}}(x,Q)\neq d_{%
\mathrm{neutron}}(x,Q)$. Experimental constraints on this effect are very
weak, even without taking into account the large uncertainties about nuclear
corrections that are needed to measure neutron structure functions \cite%
{Mrst23}. Theoretically, it has been pointed out that isospin violation in
PDFs arises naturally when one tries to include electroweak corrections in
the global QCD analysis: the evolution equations of PDFs will then include
photon distributions of the nucleon; the $u_{\mathrm{proton}}(x,Q)$ and $d_{%
\mathrm{neutron}}(x,Q)$ distributions will evolve differently due to their
different electric charge. This effect has recently been studied
\cite{Thorne}; it is small, but still can be physically relevant.

\vspace{1ex}\noindent \textbf{Heavy Quark distributions--Charm and Bottom}
\vspace{0.5 ex}%
\\Although there has been much discussion about physical processes
involving heavy quark production in the literature, there is as yet not very
much reliable
information on the heavy flavor parton distributions. Of the heavy quarks $%
c,b,$ and $t,$ only the $c$ and $b$ quark-partons actively participate in
PQCD calculations of high energy physical processes at energy
scales even up to those of LHC. The definition of heavy quark partons is even more
scheme-dependent than that of light quark flavors. In the so-called (fixed)
3-flavor scheme, there are, by definition, no heavy quark partons at all;
whereas in the (fixed) 4-flavor scheme, there is a charm distribution, but no
bottom distribution. These schemes are useful only for limited energy ranges.
In recent years, a consensus has emerged that the variable-flavor number
scheme, which is a generalization of the conventional \msbar\ zero-mass
parton scheme \cite{ACOT}, is the appropriate one to use for calculations
that cover a wide range of $Q$. If one assumes that there is no
non-perturbative heavy flavor content in the nucleon, then the heavy flavor
parton distribution functions $c$ and $b$ are ``radiatively generated'' by
QCD evolution from their respective thresholds. This is the assumption used
in all existing global analyses of PDFs. Unfortunately, the concept of
radiatively generated heavy quark partons is not fully well-defined, since
the location of ``heavy quark threshold'' for a given flavor is itself
ambiguous: it can be any value of the same order of magnitude as the heavy
quark mass or the physical heavy flavor particles. Ways to actually measure
$c$ and $b$ quark distributions at the Tevatron and LHC will be discussed in
a later section.

\vspace{-2ex}
\subsection*{Uncertainties of Parton Distributions}

\label{sec:uncert}

In parallel with the determination of ever improving ``best-fit" PDFs, an
equally important front in global analysis has been opened in recent
years---the development of quantifiable uncertainties on the PDFs and their
physical predictions. Several groups have carried out extensive studies with
different techniques and emphases. Much progress has been made; many useful
results have been obtained; but there are, as yet, no unambiguous
conclusions. The basic problem lies with the complexity of the global
analysis that utilizes results from many experiments on a variety of physical
processes, with diverse characteristics and errors, and which are often not
mutually compatible according to textbook statistics. The analyses are also
sensitive to many theoretical uncertainties that cannot be readily
quantified; and they can depend on the choice of parametrization of the
non-perturbative functions used in the analysis. Individually and
collectively, these factors render a \emph{rigorous} approach to error
analysis untenable \cite{LM,Mandy}.

As an illustration of the basic problem, we briefly describe results on a
study of the uncertainty of the W production cross section at the Tevatron
due to known experimental errors on the input data sets, conducted by the
CTEQ group \cite{cteq6} using the Lagrange Multiplier method \cite{LM}.
First, we obtain a series of PDFs that provide best fits to the global data,
but constrained to yield a series of possible values of $\sigma _{W}$ at the
Tevatron around the CTEQ6M value (which corresponds to the least overall $%
\chi ^{2}$ by definition). Then, we evaluate the individual $\chi ^{2}$ of
each experimental data set to gauge the consistency between the data sets, as
well as to assess in a sensible way to quantify the overall uncertainty of
the prediction on $\sigma _{W}$ due to the input experimental uncertainties.
The results are shown in the two plots of Fig.\,\ref{fig:wXsecErr}. %
\wXsecErr The horizontal axes correspond to the values of $\sigma _{W}.$ For
each of the 15 input experimental data sets, a best-fit value and a range is
shown. These are arranged vertically, in no particular order. On the left
plot, each range corresponds to a $\Delta\chi^2=1$ error due to that
experiment; while on the right plot, it corresponds to a ``90\% confidence
level'' (cumulative distribution function of the $\chi ^{2}$ normalized to
the best fit). We see that, if a $\Delta\chi^2=1$ error criterion is
enforced, there is no common value for the predicted $\sigma _{W}$ (or,
equivalently, some of the data sets must be deemed mutually incompatible).
But within the 90\% confidence level range, there is a common range for
$\sigma _{W}$ that spans the values indicated by the dashed vertical lines.

Faced with the problem of nominally incompatible data sets (which is common
in combined analysis of data from diverse experiments, e.g.\ PDG work),
subjective assumptions and compromise measures are necessary to obtain
sensible results. Several approaches have been followed by the different
global analysis groups. CTEQ uses the ansatz that the range of uncertainty
indicated in Fig.\,\ref{fig:wXsecErr}b represents a 90\% C.L.~uncertainty on
$\sigma _{W}$; and, in general, characterizes the PDF uncertainties by using
similar criteria along 20 orthonormal eigenvector directions in the PDF
parameter space, using an improved Hessian method.%
\footnote{%
In terms of the total $\chi ^{2}$ of the global data sets, consisting of $%
\sim 2000$ data points in current analysis, this range happens to correspond
to $\Delta \chi ^{2}\sim 100.$ There is no a priori significance to this
number, since the global $\chi ^{2}$ used in this context only represents a
broad measure of goodness-of-fit; it does not have rigorous statistical
significance. As data increase in quantity and quality, the equivalent $%
\Delta \chi ^{2}$ value will change. Similarly, when applied to a different
observable or set of input data, the value of $\Delta \chi ^{2}$ will vary.} %
MRST has adopted the same approach \cite{Mrst23}, albeit choosing a slightly
narrower range of the uncertainty. The H1 and ZEUS PDF analysis groups also
adopt similar methods, but, by restricting the input data sets to DIS
experiments only, apply their own definition of the range.~\cite{Mandy} The
important fact is that these different groups (all using the leading twist
PQCD formalism) arrive at quite comparable results, both for the PDFs and for
the magnitude of the error bands, even if some details are different because
of the variations in experimental and theoretical inputs.

With this approach, both CTEQ and MRST have been able to make estimates on
PDFs as well as their predictions on future measurements. Two examples in the
latter category are given in Fig.\,\ref{fig:Unc}. \Unc Fig.\,\ref{fig:Unc}a
shows fractional uncertainties in the predicted $q\bar{q}$ and $GG$ parton
luminosity functions as a function of $\sqrt{\hat{s}}$ at the Tevatron energy
obtained by CTEQ, from which the values and the uncertainty ranges of a
variety of physical processes, both in the SM and beyond, can be estimated.
We see a considerable uncertainty associated with the gluon-gluon luminosity
at large $x$. Fig.\,\ref{fig:Unc}b shows contours of increasing $\chi ^{2}$
in the $\sigma _{W}$-$\sigma _{H}$ plane, due to PDF uncertainties for the
LHC, obtained by MRST. Theoretical uncertainties are not included in either
plot.

A different approach is followed by Alekhin \cite{alekhin}. The experimental
input is restricted to DIS experiments only, and the theoretical framework is
broadened to include higher-twist effects, among others, in order to better
accommodate the different data sets. A consistent fit is then achieved in the
strict statistical sense; and the uncertainty range is defined according to
the classic $\Delta \chi ^{2}=1$ criterion. However, by forgoing the critical
experimental constraints provided by Drell-Yan and inclusive jet production
data, the determination of the PDFs cannot be complete. Applying the Alekhin
PDFs to the available DY data sets (E605, CDF W-asymmetry, E866), one obtains
a $\chi ^{2}$ of 892 for 145 data points---a clear indication that vital
information is missing on certain aspects of
PDFs. This can be seen in a plot of $\frac{\bar{d}-\bar{u}}{\bar{d}+\bar{u}%
}$ where the Alekhin prediction is completely different from the
experimentally determined ratio obtained from $\sigma_{pd}/\sigma_{pp}$ DY
data \cite{wktpl}. Under these circumstances, one might ask, whether these
PDF uncertainties defined by the textbook $\Delta \chi ^{2}=1$ rule are of
any practical use? Giele \textit{et al.} \cite{giele} also emphasize a
rigorous statistical approach, using the more general likelihood method.
Within the leading twist PQCD approach, this leads to acceptable results only
if one restricts the input experimental data sets to one or a few DIS sets.
Thus, depending on which subsets of data are used, one gets many predictions
on physical quantities (such as $\sigma _{W}$) with ``1$\sigma $-error''
ranges, which do not overlap with each other. This approach leaves unanswered
the important question: ``What is the best estimate of current uncertainty,
given all available experiments?''.

Thus, the underlying facts seen by all groups are consistent with one
another. The differences in interpretation lie in the emphases placed to cope
with these facts. In principle, all methods are valid and equivalent: in an
ideal world where all experiments came up with textbook-like errors, they
would all yield the same results. In reality, in the complex world of global
analysis, the results appear different or inconsistent (if strict criteria
are applied), depending on subjective judgements made in placing the
emphases. This state of affairs requires that users of PDFs must be
well-informed about the nature of the ``uncertainties'' provided by the
various global analysis groups and to use these results judiciously according
to their own (subjective) judgement and taste.

Unfortunate as it may be, but there is no ``1-$\sigma $ PDF error'' that can
be defended scientifically on all accounts. This fact points to the need for
continued hard work, both on the experimental and theoretical fronts, in
order to improve the situation, and to reduce the ambiguities described
above. To this end, the physics programs of HERA II, Tevatron Run II, as well
as several fixed-target experiments, can make important contributions in the
immediate future. Through these efforts, the PDFs and their uncertainties
will certainly be better known when the LHC comes on line. This will lead to
better predictions on both SM and new physics processes, and hence improve
the potential for all discoveries. In addition, the high reaches of the LHC,
both in energy range and in statistics, will provide additional constraints
on PDFs, and thus allow even better determination of the parton structure of
the nucleon.

\vspace{-2ex}
\subsection*{Global QCD Analysis and Collider Physics}

\subsubsection*{Standard Candle Processes and Stability of NLO QCD Predictions%
}

Because W and Z bosons are copiously produced at the Tevatron and LHC, and
because the PQCD theory for this process is well established, the W/Z cross
sections have been widely considered as ``standard candle" measurements. It
is therefore crucial to understand the uncertainty and the stability of PQCD
analyses of this process. The former has been discussed in the previous
section. We now look at the stability problem.

\vspace{1ex}\noindent\textbf{Stability of NLO global analysis and the total W
cross section}\vspace{0.5ex}%
\\
The vast majority of work on global QCD analysis of parton distribution
functions (PDFs) and the application of PDFs to the calculation of high
energy processes has been performed at NLO (1-loop hard cross sections and
2-loop evolution kernel). In recent years, some preliminary NNLO analyses
have been carried out, even if not all necessary hard cross sections are yet
available, and the evolution kernel was only known approximately (until very
recently). Since errors on most of the experimental data used in global
analysis are generally larger than the known NNLO corrections, the necessity
to extend the analysis to NNLO, at the expense of vastly more complicated
calculations, is not obvious.

A strong motivation to go to NNLO would exist, however, if the conventional
NLO analysis does not yield stable PDFs and physical predictions. This
possibility was indeed raised by a recent MRST study \cite{Mrst23}. In
particular, they found a 20\% variation in the predicted cross section for W
production at the LHC---a very important ``standard candle" process for
hadron colliders---in their NLO analysis, depending on kinematic
cuts placed on input data.\footnote{%
In the absence of general methods of assessing theoretical uncertainties in
the global analysis due to a variety of effects such as power-law corrections
(low $Q$), parton saturation (low $x$), ... etc., raising kinematic cuts on
input experimental data serves, in principle, as a poor-man's method of
reducing the theoretical uncertainties. This is done, however, at the expense
of leaving out large amounts of data that can
otherwise provide valuable constraints on the PDFs.} %
\mrst
Their results on the total cross section and on the rapidity distribution of
the W boson are shown in Fig.~\ref{fig:mrst}. For both the Tevatron Run II
and LHC physics programs this is clearly a critical issue; it is important to
investigate whether this result is confirmed by an independent study.

We have investigated this problem in considerable detail within the CTEQ
global analysis framework \cite{cteq6}. Applying the same theoretical and
experimental inputs to
the global analysis, we systematically varied the kinematic cuts in $x$ and $%
Q$, and generated new sets of best-fit PDFs. To explore the dependence of the
results on assumptions made about the parametrization of PDFs at the starting
scale $Q_{0}$ (1.3 GeV), we also studied cases where the gluon distribution
function is allow to go negative at small $x$, a possibility favored by the
MRST NLO analysis in the past few years. The main results of this study
\cite{Stability} are discussed below.

Fig.~\ref{fig:cteq}a shows the variation of the W total (W$^{+}$+W$^{-}$)
cross section at LHC with respect to a series of best fit PDFs obtained with
increasing $x$-cuts comparable to those of the MRST study; the results are
super-imposed on Fig.~\ref{fig:mrst}a. Also included is a point corresponding
to a higher $Q$-cut of $Q^{2}>10$ GeV$^{2}$.
 These results show full
stability of the predicted W cross section versus the kinematic cuts used in
the global analysis of PDFs, in marked contrast to the results of \cite%
{Mrst23}. %
\cteq%
(Results at Tevatron energies, not shown due to lack of space, are even more
stable, as one would expect.) Fig.~\ref{fig:cteq}b shows a comparison of the
predicted rapidity ($y$) distribution for (W$^{+}$+W$^{-}$) at LHC. Instead
of showing the predicted $y$-distribution from the fits with different
kinematic cuts (which are all similar to one another), we show the range of
variation of the $y$-distribution when the 2nd moment of the $y$-distribution
is extremized using the Lagrange multiplier method of \cite{LM}. These
``extreme" cases allowed by our global analysis are compared to the MRST
``default" and ``conservative" predictions. The MRST conservative prediction,
which corresponds to strong kinematic cuts, clearly stands out.

The stability of the prediction with respect to variations of kinematic cuts
seen in analysis is reassuring. It indicates that NLO QCD should provide an
adequate framework for studying high energy phenomenology at the Tevatron and
LHC, except perhaps for processes that are known to have large corrections to
the hard cross section beyond NLO.  On the other hand, the difference between
our results and those of MRST needs to be understood. The two global analysis
efforts share many common theoretical and experimental inputs. Most of their
predictions have been found to be compatible with each other. However,
historically, they also have arrived at different conclusions on a number of
specific issues, due to subtle (or not so subtle) differences in methodology
and/or input. These were eventually resolved when the causes for the
difference were identified. The case on the stability of NLO analysis may
represent such a situation. In particular, the instability of the MRST NLO
analyses appears to be associated with two other unique features of their
recent analyses: (i) an apparent ``tension" between inclusive jet production
data at relatively large $x$ and DIS data at small $x$ from HERA; and (ii) a
relatively strong preference for a negative gluon distribution $g(x,Q)$ at
small $x$ and small $Q$.

In order to look a little closer into these issues, we have investigated the
$\chi ^{2}$ values of the jet data sets and the HERA data sets separately as
we vary the kinematic cuts. No discernable improvement in the fit to the jet
data sets is seen as the $x$-cut on input DIS data is raised.%
\footnote{If there were tension between the two, an improvement would result,
because the pull of small-$x$ HERA data would have been reduced.} %
The $\chi ^{2}/N$ for the CDF and D0 experiments are $(1.47,\, 1.48,\,
1.47,\, 1.47,\, 1.48)$ and $(0.718,\, 0.715,\, 0.723,\, 0.726,\, 0.697)$ for
$x$-cuts of $(0,\, .001,\, .0025,\, .005,\, .01)$ respectively. The quality
of fit to the HERA data (as measured by $\chi ^{2}/N$) also remains stable
with respect to the change in the kinematic cuts.

An important effect of raising the kinematic cuts in $x$ and $Q$ is that
constraints provided by the precision HERA data in the small $x$ and low $Q$
region are removed from the global fit. Thus, the uncertainty of the
resulting PDFs and their physical predictions will increase. We quantified
this effect for the specific case of W production at LHC. Using the Lagrange
Multiplier method \cite{LM}, we examine the best $\chi ^{2}$ values obtained
in constrained global fits, as a function of the W cross section,
for several choices of kinematic cuts: (i) the \emph{normal} CTEQ cuts ($%
Q^{2}>4$ GeV$^{2}$); (ii) \emph{medium} cuts ($Q^{2}>6.25$ GeV$%
^{2},~x>0.001$); (iii) \emph{strong} cuts ($Q^{2}>10$ GeV$^{2},~x>0.005$);
and (iv) \emph{very strong} cuts ($Q^{2}>100$ GeV$^{2}$).\footnote{%
A $W$-cut of $3.5$ GeV is maintained in all cases. Note that the strong cuts
are similar to those of the MRST conservative fit.} Fig.~\ref{fig:parabolas}a
shows the results when the normal CTEQ parametrization, with positive PDFs at
the input scale of $Q=1.3$ GeV, is used. We see that, as stronger kinematic
cuts are imposed, the range of uncertainty on the prediction for the W
production cross section becomes progressively larger. Whereas the change
from the normal to medium cuts is small, the range of uncertainty almost
doubles from normal to strong cuts at any tolerance level of $\chi^2$. The
\emph{very-strong-cuts} case removes most DIS data from the fit, thus heavily
emphasizes collider data---it represents a step toward a hypothetical
``collider-data-only prediction" from Tevatron to LHC. It results in a very
large range of uncertainty for the LHC cross section, as one would have
expected.

\parabolas

We have also performed similar fits, but removing our usual constraint that
parton distributions are positive definite at the scale $Q_0$. This affects
mostly the gluon, as quark distributions are more directly
related to (positive) input structure functions.%
\footnote{A strong enough negative initial gluon at small $x$ can induce
negative quark distributions at small $x$ and larger $Q$ through QCD
evolution.  This happens to the MRST conservative PDF set, for instance,
at $Q^2=5 $ GeV$^2$ for $x<0.00015$.}%
 As a rule, our best fits for any given sets of kinematic cuts do
not require negative gluons at our input scale of $Q_0=1.3$ GeV. However,
when $\sigma_W^{LHC}$ is pulled away from the best-fit value by the
constrained fits in the Lagrange Multiplier study, the additional freedom
provided by a negative gluon widens the allowed range of variation of
$\sigma_W^{LHC}$. The results of this analysis (except that of the very
strong cuts) are shown in Fig.~\ref{fig:parabolas}b. Compared to the
corresponding cases in Fig.~\ref{fig:parabolas}a, we see that the medium-cuts
curve has opened up noticeably at the lower end, and the strong-cuts case has
widened at both ends.

Should the fits with negative gluons be taken seriously as candidate PDF
sets, hence providing better determination of the true range of uncertainty?
We think not. The reason is that whereas PDFs are not strictly forbidden to
become negative, \emph{\textbf{all} physical quantities calculated from them,
in \textbf{all} parts of physical phase space, must remain positive
definite}. This is a very strong condition that is extremely difficult to
satisfy if some PDFs become negative in some region of the $(x,\,Q)$ plane.
For all the solutions involving negative gluons at small $x$ and low $Q$
(including the MRST ones), we found it is possible to identify
\emph{\textbf{some}} physical cross sections, at \emph{\textbf{some}} high
energies, near \emph{\textbf{some}} boundary of phase space, that become
negative. These negative gluon solutions must therefore be considered as
unphysical (at this order of $\alpha_s$).

Another possible source of discrepancy between the CTEQ and MRST results is
the treatment of heavy quark mass effects in DIS. We primarily use the widely
used zero-mass \msbar\ formalism, whereas MRST uses the Thorne-Roberts
prescription for the non-zero-charm-mass variable flavor number scheme. In
general, the difference is nominally small. However, since charm production
accounts for nearly 25\% of the DIS cross section at small $x$, and since the
HERA data at small $x$ have very small errors, the difference could have an
effect on the issues discussed above, because the R-T prescription results in
an unusual behavior of the charm structure function
just above the threshold. %
Further study is clearly needed to settle fully the stability issue.

We should add that, independent of the stability of NLO global analysis, NNLO
calculations are needed for processes that require a high level of accuracy,
or that are known to have large corrections.

\vspace{-2ex}
\subsubsection*{Precision W/Z Differential Distributions as Input to PDF
Analysis}

The differential cross section for W/Z production $d^2\sigma \,/\,dy\,dp_{T}$
(or, more practically, the cross section $d^2\sigma \,/\,dy\,dp_{T}$ for one
of the decay leptons in the semi-leptonic decay channel) is sensitive to
details of PDFs. Precise data on these cross sections can play a decisive
role in narrowing the uncertainties and clarifying many of the open issues on
PDFs described in the first part of this review. This is because: (i) they
measure completely different combinations of PDFs, thus provide constraints
on many independent quantities not accessible in DIS experiments; and (ii)
the kinematic coverage of the collider cross section data will greatly expand
that of available DIS data. It is particularly important that the W/Z cross
sections be measured as precisely, and in as wide a kinematic range, as is
possible at the Tevatron, in order to determine the PDFs well enough to
enable better predictions, hence improved discovery potentials, at the LHC.

The Tevatron and LHC are \emph{W/Z factories}. The reason that their
potential for contributing to the next generation of global QCD analysis has
so far not attracted much attention has perhaps to do with the fact that the
measured cross sections, involving convolutions of the products of two PDFs,
do not depend on the PDFs in as direct a way as the structure functions of
DIS scattering. Thus, it is difficult to highlight which measurement will
determine what particular features of PDFs. But, since most of the open
issues in current PDF analysis concern subdominant effects, the more subtle
role to be played by precision W/Z data will be both natural and vital.
Instead of looking at LO parton formulas for motivation to focus on certain
measurements, we now need detailed phenomenological studies of the effects of
various measurements on the remaining uncertainties of PDFs in the global
analysis context, utilizing the new tools developed in recent years, such as
the Lagrange multiplier method. Efforts of this kind have not yet been
systematically carried out; but are crucial for the success of the Tevatron
and LHC physics programs.

Many of the same comments apply to cross sections of W/Z plus jets, for which
there will also be abundant data. A great deal of theoretical work has been
done recently on this subject. Since the definition of jets, and related
issues also come into play, the detailed study of this process will probably
concern less about PDF analysis, and a lot more with new physics discovery
backgrounds.

\vspace{-2ex}
\subsubsection*{W Mass Measurement}

The precise determination of the W mass is one of the key measurements at
hadron colliders. One of the main uncertainties of this measurement is the
error attributable to PDFs. In spite of a great of deal of effort, current
estimates are still not well founded. Detailed studies of the kind described
in the previous section can help expose the effects due to uncertainties in
regions of the PDF parameter space that have not been included in previous
estimates, as well as identify new measurements, prior to the full mass
measurement analysis, that can reduce the uncertainties. A concerted effort
by theorists and experimentalists working together is essential in this
endeavor, because the task involves a strong interplay between theoretical
and experimental considerations.

\vspace{-2ex}
\subsubsection*{W/Z Plus Tagged Heavy Flavor Production and Heavy Quark
Distributions}

One area where hadron collider measurements can, in principle, make clearly
defined contributions to the parton structure of the nucleon is heavy quark
distributions, $c(x),b(x)$. The relevant processes involve the production of
a W/Z boson with an
associated heavy flavor meson, as illustrated here:\vspace{1ex}
\\%
\centerline{
 \resizebox{0.8\textwidth}{!}{
  \includegraphics[clip=]{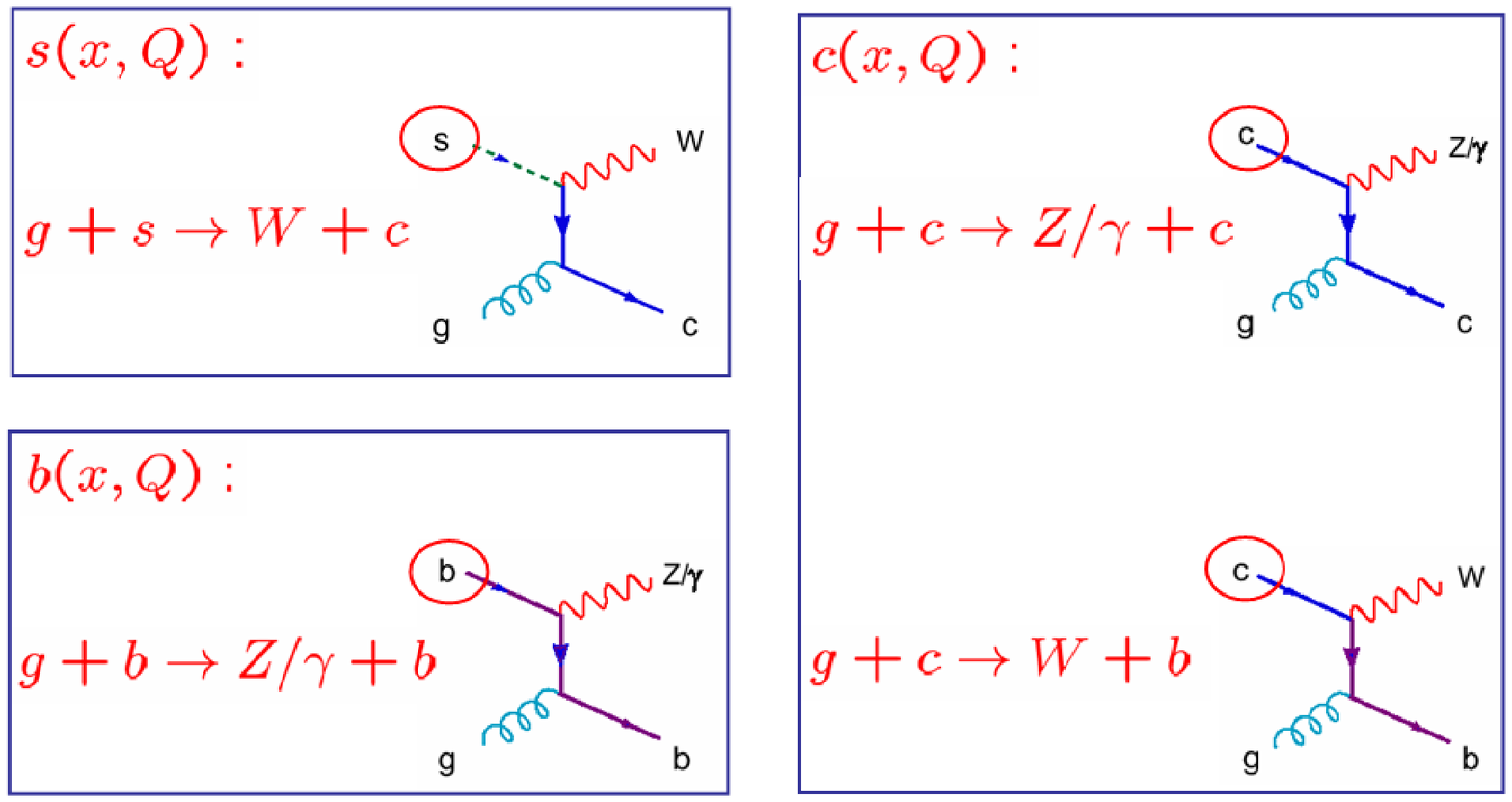}
 }
}\\%
These final states are very difficult to measure reliably and precisely. But
they do provide direct handles on the heavy quark distributions of the
nucleon that are not available otherwise. Hence, it is a very important
challenge that needs to be met.

\vspace{-2ex}
\subsubsection*{Predictions for Top Production at Hadron Colliders}

The quantitative study of the top quark properties and its production cross
sections is a high priority both at the Tevatron and at LHC. The
uncertainties of PQCD predictions on the $t\bar{t}$ production cross section have
been studied extensively. As an example, some of the results of the detailed
study of \cite{ttbar} are captured in the following table.\\%
\centerline{
 \resizebox{0.75\textwidth}{!}{
  \includegraphics[bb=60 221 780 649,clip=]{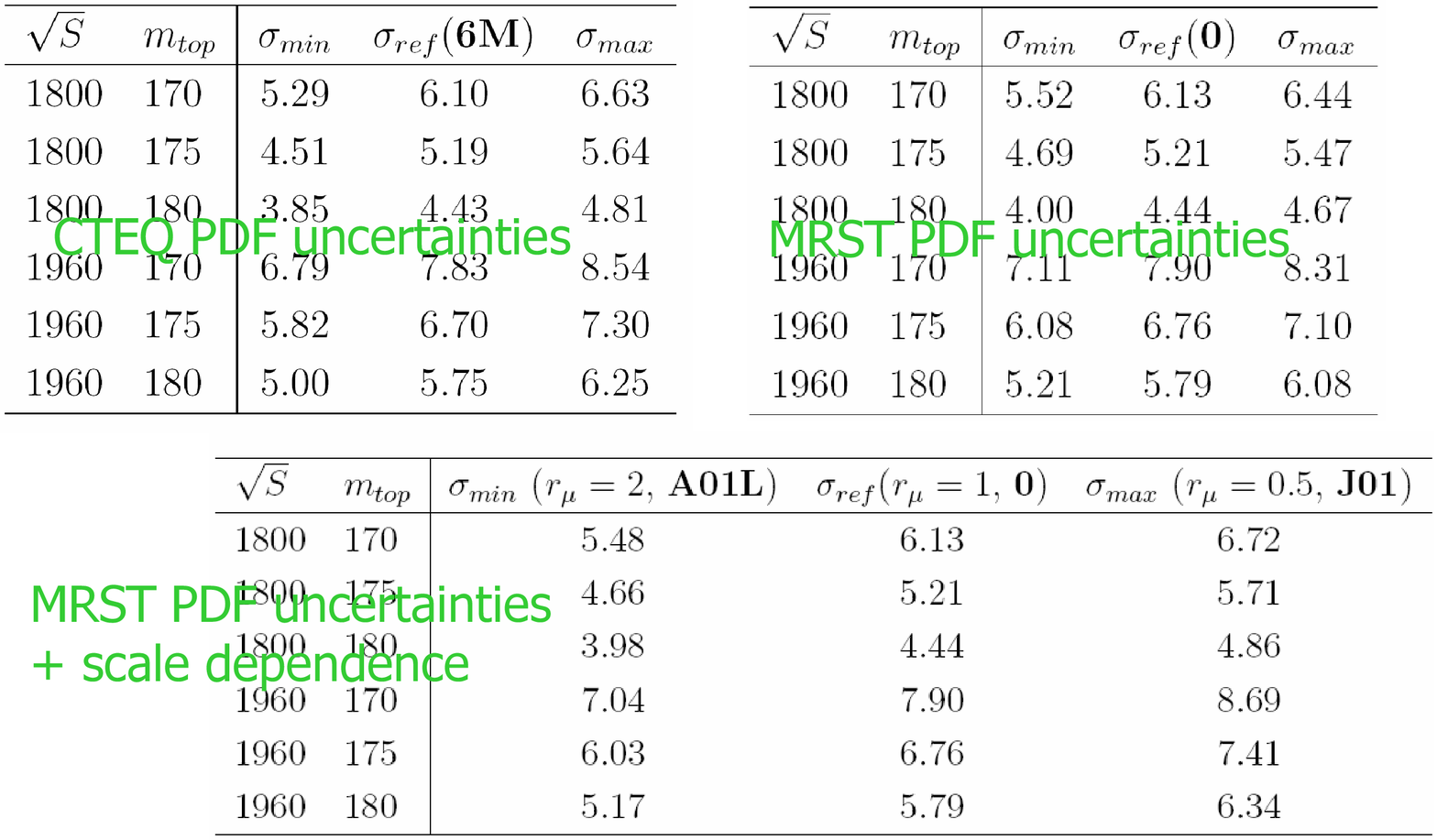}
 }
}\\%
The first row shows the range of predicted cross sections associated with
uncertainties of PDFs due to experimental inputs, as given by CTEQ and MRST.
The second row shows the range due to the variation in $\alpha _{s}$, with
associated change in PDFs, as given by MRST.

The anticipated measurement of single top production will reveal valuable
information on the top coupling to gauge bosons, among other things. The
production mechanism, involving an intricate interplay between $%
gW\rightarrow \bar{b}t$ and $bW\rightarrow t$ subprocesses, provides an
important forum to study the PQCD physics of heavy quark partons. Although
the basic theory behind this physics has, by now, been well-established \cite%
{ACOT,collins}, there are many implementation issues, as well as the general
lack of information on the $b$ distribution (cf.~previous section on heavy quark
PDFs), that make this process an extremely interesting one to study at both
hadron colliders.

\vspace{-2ex}
\subsubsection*{Predictions for Higgs Production at Hadron Colliders}
The intense interest in discovering the Higgs particle, whether within the SM
or beyond, also underlines the importance of having precise knowledge on
PDFs. The many relevant issues will be explored in detail in other sessions
of this workshop. We will only quote one example: the uncertainty of the
predicted cross section due to the subprocesses $gg\rightarrow HX$ \ is
clearly tied to the (rather large) uncertainty on the gluon distribution,
which we discussed earlier. This uncertainty is larger at the Tevatron than
at LHC, mainly because of the different $x$-range involved. The results of
\cite{DjFe}, comparing the predictions of three recent PDFs, are shown below.\\%
\centerline{
 \resizebox{0.75\textwidth}{!}{
  \includegraphics[clip=]{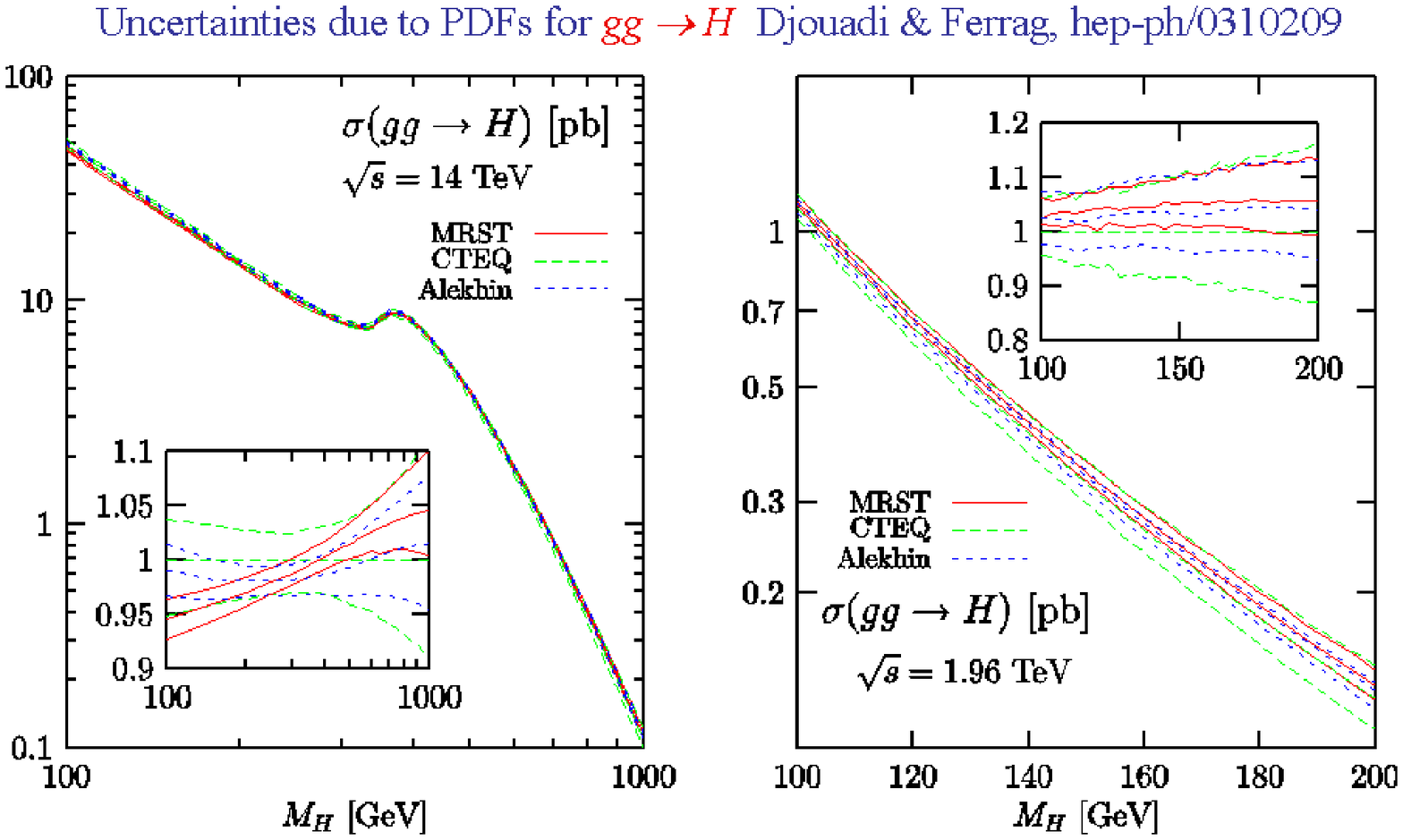}
 }
}\\%

\vspace{-6ex}
\subsection*{Conclusions}

The importance of precise knowledge of PDFs for calculating SM processes and
making predictions of New Physics discoveries at the hadron colliders has
been well recognized. In reviewing this subject, we emphasize that, in spite
of the steady progress made in global QCD analysis, there are still many gaps
in our understanding of the parton structure of the nucleon. Continued
advances to fill these gaps will require a sustained effort by the whole HEP
community.

There are many measurements at the Tevatron and LHC that can help. For
instance:%
\mylis[$\bullet$]%
{%
\item many processes at the high energy colliders are
dominated by gluon initiated subprocesses. Better data on these processes
(such as jet cross sections, heavy quark production, ... etc.) will provide
much needed constraints on the not-so-well-determined gluon distribution;

\item since the colliders are W/Z factories, precisely measured rapidity and
transverse momentum distributions can provide constraints on the quark
distributions that are complementary to those afforded by DIS data. For
instance, the W$^\pm$ rapidity distributions at the LHC can differentiate between
$u(x)$ and $d(x)$ without the uncertainties of nuclear corrections that beset
current analyses that use deuteron targets;

\item W/Z plus heavy flavor channels provide unique handles on heavy quark
distributions, which are, so far, virtually untested. %
}%

\noindent All these can have significant impact on the precision measurement
of the W mass and top quark parameters, as well as on the discovery
potentials for Higgs and New Physics signals.


\begin{thebibliography}{99}

\bibitem{Mrst23}
MRST: A.~D.~Martin, \emph{et al.},
Eur.\ Phys.\ J.\ C {\bf 28}, 455 (2003);
Eur.\ Phys.\ J.\ C \textbf{35}, 325 (2004).
%

\bibitem{cteq6} CTEQ: J.~Pumplin, \emph{et al.},
JHEP \textbf{0207}, 012 (2002);
%
D.~Stump, \emph{et al.},
JHEP {\bf 0310}, 046 (2003).

\bibitem{NuSea} J.~C.~Webb \textit{et al.} [NuSea Collaboration],
hep-ex/0302019; D. Isenhower, Proceedings of DIS2004.

\bibitem{CN} CCFR and NuTeV Collab.~(M.~Goncharov \textit{et al.}),
Phys.~Rev.~\textbf{D64}, 112006 (2001); NuTeV Collab.~(M.~Tzanov \textit{et
al.}), \texttt{hep-ex/0306035}; P.~Spentzouris, Proceedings of DIS2004.


\bibitem{NuTeV} NuTeV Collaboration, G.P.~Zeller \textit{et al.},
Phys.~Rev.Lett.~\textbf{88}, 091802 (2002); for physics consequences, see,
e.g.\ S.~Davidson, \emph{et al.}, JHEP \textbf{%
0202}, 037, 2002.

\bibitem{StrAsym} F.~Olness \textit{et al.}, hep-ph/0312323;
S.~Kretzer \textit{et al.},
Phys.\ Rev.\ Lett.\ \textbf{93}, 041802 (2004); S.~Kretzer, Proceedings of DIS2004,
hep-ph/0408287. 

\bibitem{Thorne}
R.~S.~Thorne, Proceedings of DIS2004, hep-ph/0407311.
%
\bibitem{ACOT}
ACOT: M.~Aivazis, \emph{et al.},
Phys.\ Rev.\ D {\bf 50}, 3102 (1994);

\bibitem{LM} D.~Stump \textit{et al.},
Phys.\ Rev.\ D \textbf{65}, 014012 (2002); 
J.~Pumplin \textit{et al.},
Phys.\ Rev.\ D \textbf{65}, 014013 (2002); and Phys.\ Rev.\ D \textbf{65},
014011 (2002). 

\bibitem{Mandy}
A.~M.~Cooper-Sarkar,
J.\ Phys.\ G {\bf 28}, 2669 (2002) [arXiv:hep-ph/0205153];
\\
also http://www-zeuthen.desy.de/~moch/heralhc/gwenlan-coopersarkar.pdf

\bibitem{alekhin}
S.~Alekhin,
Phys.\ Rev.\ D {\bf 68}, 014002 (2003), and references therein.
%
\bibitem{wktpl} W.K.\ Tung, plenary talk, Proceedings of DIS2004,
hep-ph/0409145. 

\bibitem{giele}
W.~T.~Giele and S.~Keller,
Phys.\ Rev.\ D {\bf 58}, 094023 (1998);
%
W.~T.~Giele, S.~A.~Keller and D.~A.~Kosower,
hep-ph/0104052 (unpublished).
%
\bibitem{Stability} J.~Huston, J.~Pumplin, \textit{et al}, under preparation.

\bibitem{ttbar}
M.~Cacciari, \emph{et al.},
JHEP {\bf 0404}, 068 (2004).
%
\bibitem{collins}
J.~C.~Collins,
Phys.\ Rev.\ D {\bf 58}, 094002 (1998) [arXiv:hep-ph/9806259].
%
\bibitem{DjFe}
A.~Djouadi and S.~Ferrag,
Phys.\ Lett.\ B {\bf 586}, 345 (2004).


\end{thebibliography}
\end{document}